\documentstyle[11pt,Cargese,twoside,epsf]{article}
\def\edcomment#1{\iffalse\marginpar{\raggedright\sl#1\/}\else\relax\fi}
\reversemarginpar

\begin{document}
\title{MBM12: A Younger Version of the TW Hydrae Association?}
 \author{Scott J. Wolk }
\affil{Harvard--Smithsonian Center for Astrophysics, Cambridge,
MA 02138, U.S.A.}
\author{Rebecca T. Cover}
\affil{Williams College, Williams, MA., U.S.A.}
\author{Ray Jayawardhana}
\affil{Dept. of Astronomy, University of California, Berkeley, CA 94720,
U.S.A.}
\author{Thomas J. Hearty}
\affil{Jet Propulsion Laboratory, Pasadena, CA., U.S.A. }

\begin{abstract}
We present optical and infrared photometry of stars near the MBM~12
dark cloud.  The optical data indicate that there are about 40
candidate pre--main sequence stars in the same area currently
occupied by eight bona--fide PMS stars. Analysis of 2MASS data suggests
that 60\% of PMS candidates had been detected. Some of these show
evidence of optically thick disks. Our study of the 2MASS data also 
reveals a large enhancement of sources within the MBM~12 cloud, 
indicating that it is still actively forming stars.

\end{abstract}

\section{Introduction}

MBM~12 is a dark, high latitude cloud located 
about 65 pc from the sun (Hobbs, Blitz \& Magnani 1986, 
but see Hearty et al. 2000).  From the 
early 1970s through  early 1990s, five Classical T Tauri stars (cTTs) and
three naked T Tauri stars (nTTs) were discovered associated with the
cloud (see Hearty et al. 1999 and references therein). 
These sources have been discovered through a combination of H$\alpha$
objective prism surveys and X-ray surveys.
Such a close,
young, isolated association would be in a league with the TW Hydrae
Association (Kastner et al.~1997) and Eta Cha (Mamajek et al. 1999 and
Lawson et al. this volume).  However, the co--existence of the cloud
and several stars with disks implies this is a 
very young association of stars which are very close
to the earth. The 10-Myr-old TW Hydrae group has proved a fruitful
region for high resolution imaging of disks and close binary companions
(Jayawardhana 2000). At an estimated age of about 1 Myr, in MBM~12 
brown dwarfs/Jupiters would be brighter and disks should be more numerous.

As a  close, possibly active region of star formation, still nestled near
its natal cloud,  MBM~12 vitally requires a census of its young stellar
population . Here, we present the results of our study of the photometric
properties of stars in this region.

\section{Optical Photometry}

\begin{figure}
\plotone{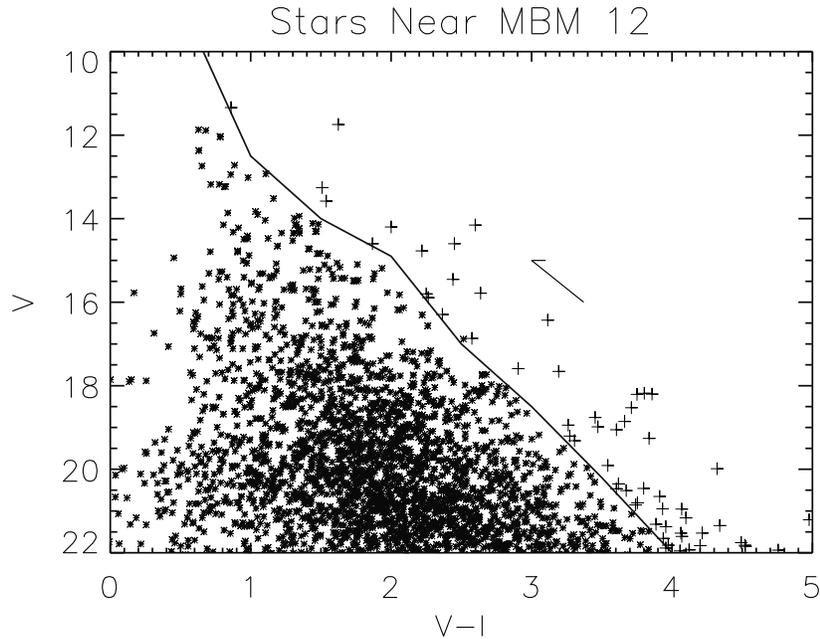}
\caption{An optical color--magnitude diagram for the observed regions
near MBM~12. The crooked line indicates the location of the Pre--main
sequence for 10-Myr-old stars at 65 pc (from D'Antona \& Mazzitelli,
1997).  Stars above this line
are marked with plus signs.  The arrow indicates 1 A$_v$ of reddening.}
\end{figure}

We observed the stars near the cloud in optical light during December
1998 with the 48$^{\prime\prime}$   telescope at the 
Whipple Observatory using the 
``four-shooter'' CCD detector.
There were a total of 6 pointings made, each centered on a known
T Tauri star.  The field of view for each image was about
13.5$^{\prime}$.  
Five of the pointings coincided with the edges of the MBM~12 cloud.  
The sixth was 2$^{\circ}$ away, closer to MBM~13.
Observations were made in U, B, V, R and I.  There was a series of short
60 second V, R and I exposures, along with a deeper 120 to 450 second
exposure in all five Kron-Cousin's bands.

The data were corrected in the usual way using IRAF. The only
complication was that the ``four--shooter'' uses four separated CCDs
for data taking.   For processing, the data from the individual chips 
were segregated. 
Aperture photometry was performed on each night's worth of data on a
per chip basis.  This way, each chip had its own set of standard stars
so that effects induced by a chip would be confined
to that data set.  Cross--referencing of stars observed on multiple
chips showed chip to chip variations of less than 1\%. 
The overall photometric accuracy was about 3\% fainter than V=21. The data
are complete to about V=21 and I =17.5.

The main results of these observations are shown in figure~1.  About
10,000 stars were detected in V, R and I bands. Brighter than V=21, 
all but about 40 of these
stars fall well below the demarcation for 10 Myr old stars at 65pc.
We define these as our PMS candidates. % Note, we do not believe all of
%these stars are PMS, we aare especially suspicious of the group below
%V=20, within 0.5 mag. of the cutoff in V$-$I.
There is a small ($\sim$0.5 mag.) break between the main body 
of stars and the PMS candidates.  This could be the effect of
clustering but is most likely an effect of volume limiting.  The
volume of space a star could be in, and occupy a given location on the
color--magnitude diagram becomes relatively small this close to the
sun. 

Spatially the candidates are spread among the 6 fields fairly evenly.
The field near MBM~13 does have fewer PMS candidates than the other 5
fields, but this number is still within 2$\sigma$ of the mean.  We do not 
see any candidates along the edge of the cloud.  This
could be due to high extinction. But a very small fraction of the
observed regions (only about 5\%) overlaps the MBM~12 cloud.  So this
result could simply be happenstance.

\section{2MASS Photometry}

\begin{figure}
%\plottwo{mbm12_cand_ccd.ps}{IRDENSITY.ps}
%\plotone{mbm12_cand_ccd.ps}
\plotfiddle{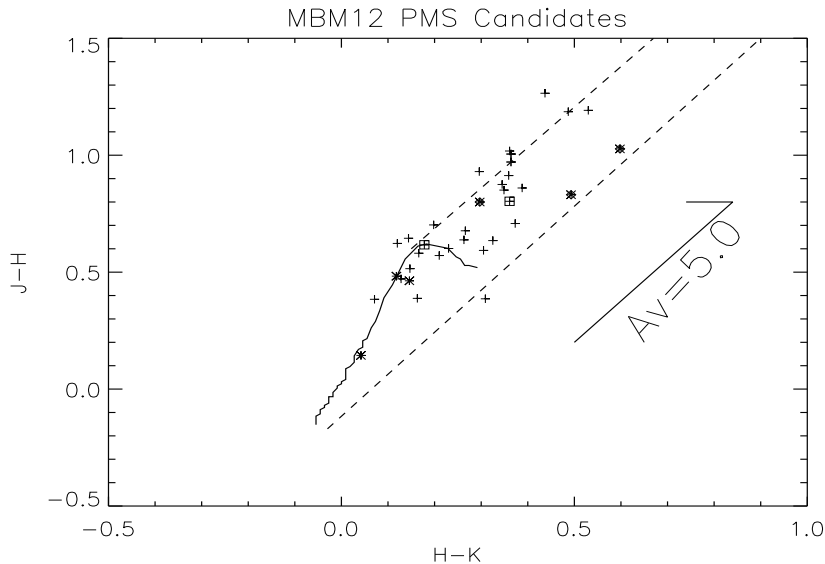}{2.6in}{90}{55}{50}{200}{-62}
\caption{An infrared color--color diagram, the plus signs indicate
optically selected PMS candidates near MBM~12 from figure~1.   
Stars marked with diamonds are known cTTs. Stars marked with
squares are known nTTs.  Stars from the HIPPARCHOS catalog are 
marked with asterisks and are used for reference. A 5 visual magnitude
reddening vector is indicated.}
\end{figure}

\begin{figure}
%\plotfiddle{file}{vsize}{rot}{hsf}{vsf}{htrans}{vtrans}
%\plotone{IRDENSITY.ps}
\plotfiddle{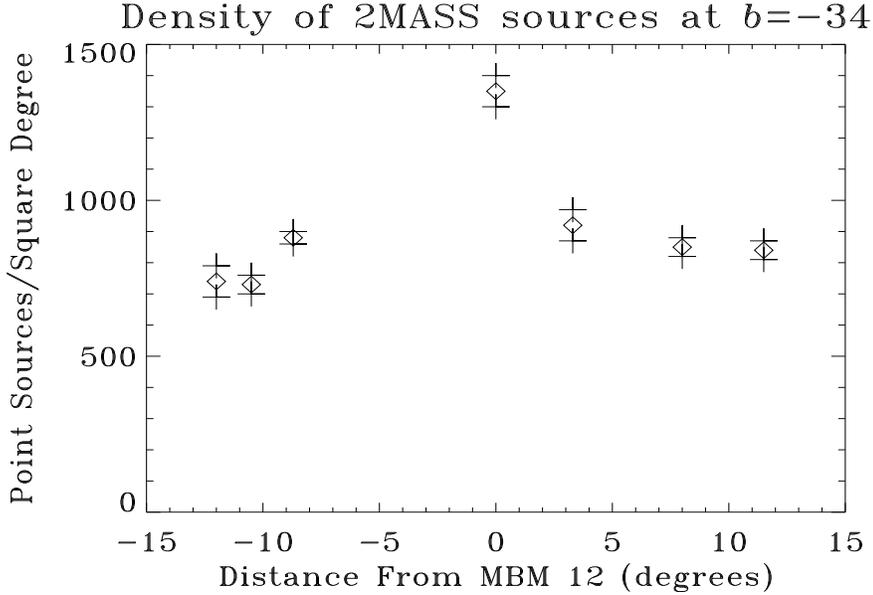}{2.9in}{90}{55}{50}{200}{-50}
%\caption{A plot of the density of 2MASS sources as a function of
%position relative to the MBM12 cloud.  All measurements are made within
%a degree of the same galactic latitude. The density of sources is
%fairly steady at 800 stars/square degree except for at the center of the MBM12
%cloud.}
\caption {A plot of the density of 2MASS sources as a function of
position relative to the MBM~12 cloud.  All measurements are made within
a degree of the same galactic latitude. The density of sources is
fairly steady at 800 stars/square degree except for the center of the MBM~12
cloud.}
\end{figure}

We obtained several nights to survey MBM~12 from the Whipple Observatory
using the STELLIRCAM near--infrared camera. Unfortunately the weather
did not cooperate.  Fortunately, the second incremental release of the 
2MASS database covers the MBM~12 region.  We queried the
database in the region of the optical surveys (within 3600 arcsec
radius of  RA:3h57m DEC:20$^o$15$^\prime$ J2000).  This returned about
5500 sources.  We then filtered out all the sources with errors
greater than 10\% or any bad quality flags.
%either source confusion or artifact related. 
This left a little under 4000 sources with which we  
correlated with our catalog of $\sim$40 PMS candidates 
looking for matches within 3$^{\prime\prime}$.  
We found 24 matches.  To this we added the
previously known TTs in this region; these stars were too bright to be
measured with our optical photometry. Six of these 8 were in the
2MASS catalog. We also used 3 HIPPARCHOS stars as reference for normal
stars along the line of sight. 

The results are given in figure~2. Two stars are found in the
lower--right region
of the diagram, which indicates the existence of circumstellar disks.
About half of the PMS candidates are
found in the region of the diagram which indicates a normal
photosphere with significant (A$_V~>1$) reddening. Being found in this
region does not mean
that these stars do not have disks; in fact the four known cTTs are
found in this region as well.
Nor does the substantial reddening found in the IR 
mean that the optical data are misleading.  The reddening vector
in figure~1 is parallel to the (pre--) main sequence, so reddening
only affects the mass estimate.  For this reason, we refrain from
attempting any sort of mass function at this point.
The remaining stars lie close to the main sequence locus.

The large sky coverage of the 2MASS database allowed us to 
study the spatial distribution of sources.  Using the second
incremental release point source catalog, we created pencil beam
samples of the infrared sky at various locations along the same latitude
as MBM~12.  Each pencil beam was 20$^{\prime\prime}$ in radius and so
covered 0.35$^{\circ 2}$.
One pencil beam was centered on the MBM~12 cloud itself, one on the field
near MBM~13, and others were spread further out, but always on areas
with warm dust as seen by the IRAS 60\micron~ survey.  Figure~3 shows
that the pencil beam 
centered on the MBM~12 dark cloud has a much greater density of 
sources than the
other regions.  Using a 20$^{\prime\prime}$ pencil beam, the enhancement
along MBM~12 is about 10$\sigma$.  However, this result becomes even
more significant if smaller pencil beams are used.  The total
enhancement is about 70 objects more than expected in the
20$^{\prime\prime}$ pencil beam.

\section{Conclusions}
The proximity and extreme youth of stars near the MBM~12 cloud makes
them of great interest.  While the known membership of the cluster is
small, we have found evidence for dozens of additional cluster members
near the cloud and perhaps 70 embedded within the cloud.  Over the
coming observing seasons, we hope to be able to follow up on
this photometric survey with spectra which will confirm the PMS nature
of the candidates.  These new data will help us better ascertain the 
number and age of these nearby stars.

\acknowledgments 
This work was supported in part by NASA contract NAS8-39073.


\begin{references}
D'Antona, F.\& Mazzitelli, I. 1997 Mem. S.A.It., 68, 807\\
Hearty, T. et al. 2000, A\&A, 357, 681\\%DISTANCE
Hearty, T. et al. 2000, A\&A, 353, 1044\\%XRAY
Hobbs, L.M. et al. 1986, ApJ, 306, 380L\\ 
Jayawardhana, R.  2000, Science, 288, 64\\
Kastner, J.H.,  et al. 1997, Science, 277, 67 \\
Mamajek et al. 1999, ApJ, 516, L77\\  
Lawson et al. 2000, this volume\\
\end{references}
\end{document}